# Quantum Entanglement and the Two-Photon Stokes Parameters

Ayman F. Abouraddy, Alexander V. Sergienko[†], Bahaa E. A. Saleh, and Malvin C. Teich

*Quantum Imaging Laboratory,[‡] Departments of Electrical & Computer Engineering and Physics, Boston University, 8 Saint Mary's Street, Boston, Massachusetts 02215-2421*

## Abstract

A formalism for two-photon Stokes parameters is introduced to describe the polarization entanglement of photon pairs. This leads to the definition of a *degree of two-photon polarization*, which describes the extent to which the two photons act as a pair and not as two independent photons. This pair-wise polarization is complementary to the degree of polarization of the individual photons. The approach provided here has a number of advantages over the density matrix formalism: it allows the one- and two-photon features of the state to be separated and offers a visualization of the mixedness of the state of polarization.

---

[†] Electronic address: alexserg@bu.edu

[‡] URL: http://www.bu.edu/qil

In his paper of 1852 [1], G. G. Stokes studied the properties of beams of light in an arbitrary state of polarization and devised four parameters, known since as the Stokes parameters [2], which completely specify the polarization properties of a beam of light. The Stokes parameters have been an essential element in the development of various metrological techniques that involve the use of polarized light. These parameters have recently been extended to the quantum domain [3].

The development of new nonclassical sources of light exhibiting polarization entanglement demands that the Stokes parameters be further extended. In this paper we introduce a generalization of the Stokes parameters to two-photon sources, referred to hereinafter as 2P-SP. Although the density matrix provides a complete description of the state, the formalism presented here is advantageous for characterizing the unusual properties of such sources from conceptual and computational points of view. The properties of the two photons as individuals, versus their properties as a pair, can be more readily distinguished in the proposed formalism.

Consider a source emitting two photons, as shown in Fig. 1. The implementation of such a source is readily achieved via spontaneous parametric down-conversion in a second-order nonlinear crystal pumped by a laser beam [4]. Such sources have become an essential ingredient in many experimental realizations of the new field of quantum information processing [5], and have also been used in the emerging field of quantum metrology [6].

*One-photon Stokes parameters.*— Of the many definitions of the Stokes parameters, the one that suits our purpose best is the definition noted in the early days of quantum mechanics [7], where the Stokes parameters, $S_j, j = 0,..,3$, are the (real) coefficients of expansion of the 2×2 polarization density matrix $\boldsymbol{\rho}_1$ in terms of the Pauli matrices $\boldsymbol{\sigma}_j$, $j = 0,..,3$ [8]:

$$\boldsymbol{\rho}_1 = \frac{1}{2}\sum_{j=0}^{3} S_j \boldsymbol{\sigma}_j, \quad S_i = \text{Tr}(\boldsymbol{\sigma}_i \boldsymbol{\rho}_1), \quad S_0 = \text{Tr}(\boldsymbol{\rho}_1) = 1. \tag{1}$$

The state of polarization of a one-photon source can also be described by the Stokes parameters, referred to hereafter by 1P-SP. One can use the 1P-SP to distinguish between two classes of such sources, pure- and mixed-state beams, via a one number that is a function of the 1P-SP, namely the degree of polarization defined as $P_1 = \sqrt{S_1^2 + S_2^2 + S_3^2}$ [2]. A pure state of polarization yields $P_1 = 1$. A maximally mixed (unpolarized) state yields $P_1 = 0$, so that the 1P-SP are $\{1,0,0,0\}$.

*Two-photon Stokes parameters.* — For the case of a two-photon source we extend the definition in Eq. (1), defining the 2P-SP as the (real) coefficients of expansion of the 4×4 polarization density matrix $\boldsymbol{\rho}_{12}$ of the photon pair in terms of *two-photon Pauli matrices*, $\boldsymbol{\sigma}_{ij} = \boldsymbol{\sigma}_i \otimes \boldsymbol{\sigma}_j$, $i, j = 0,..,3$:

$$\boldsymbol{\rho}_{12} = \frac{1}{4}\sum_{i,j=0}^{3} S_{ij} \boldsymbol{\sigma}_{ij}, \quad S_{ij} = \text{Tr}(\boldsymbol{\sigma}_{ij} \boldsymbol{\rho}_{12}), \quad S_{00} = \text{Tr}(\boldsymbol{\rho}_{12}) = 1. \tag{2}$$

We now have a set of 16 2P-SP [9]. The 16 two-photon Pauli matrices are linearly independent and can be used as a basis for the linear vector space of 4×4 matrices defined over the field of complex numbers.

An important feature of the definition of the 2P-SP is that the 1P-SP for each photon are included within them as a subset. The reduced density matrix of the first photon (after tracing over the subspace of the other photon in $\boldsymbol{\rho}_{12}$) is

$$\boldsymbol{\rho}_1 = \begin{pmatrix} \rho_{11} + \rho_{22} & \rho_{13} + \rho_{24} \\ \rho_{13}^* + \rho_{24}^* & \rho_{33} + \rho_{44} \end{pmatrix}, \tag{3}$$

where the elements indicated are those associated with $\boldsymbol{\rho}_{12}$, so that the 1P-SP are

$$\mathbf{S}_1 = \begin{bmatrix} S_0 \\ S_1 \\ S_2 \\ S_3 \end{bmatrix} = \begin{bmatrix} \rho_{11} + \rho_{22} + \rho_{33} + \rho_{44} \\ \rho_{11} + \rho_{22} - \rho_{33} - \rho_{44} \\ 2\,\mathrm{Re}\,\rho_{13} + 2\,\mathrm{Re}\,\rho_{24} \\ 2\,\mathrm{Im}\,\rho_{13} + 2\,\mathrm{Im}\,\rho_{24} \end{bmatrix} = \begin{bmatrix} S_{00} \\ S_{10} \\ S_{20} \\ S_{30} \end{bmatrix}, \tag{4}$$

and similarly for $\mathbf{S}_2$. All the 2P-SP with 0 in their index thus represent the 1P-SP for the two photons separately.

*Measurement of the two-photon Stokes parameters.* — The question arises of how to measure the 2P-SP. Referring to the configuration illustrated in Fig. 1, polarization measurements may be carried out on each beam separately (singles measurements), or carried out simultaneously on the two beams (coincidence measurements). In general it is not possible to characterize the state of polarization by ones measurements alone, since the state may be entangled [10]. There are many schemes for performing such measurements [11], one of which relies on carrying out only coincidence measurements with various polarization analyzers placed in the two beams. An example of a set of 16 sufficient coincidence measurements is provided by:

→⊗→, →⊗↗, →⊗↻, →⊗↺, ↗⊗→, ↗⊗↗, ↗⊗↻, ↗⊗↺, ↻⊗→, ↻⊗↗, ↻⊗↻,

↻⊗↺, ↺⊗→, ↺⊗↗, ↺⊗↻, ↺⊗↺,

where → and ↗ are horizontal and 45° linear polarization analyzers (with respect to some chosen direction), respectively; and ↻ and ↺ are right-hand and left-hand circular polarization analyzers, respectively. Each analyzer may be expressed in terms of Pauli matrices as follows: $\rightarrow = \frac{1}{2}(\sigma_0 + \sigma_1)$, $\nearrow = \frac{1}{2}(\sigma_0 + \sigma_2)$, $\circlearrowright = \frac{1}{2}(\sigma_0 - \sigma_3)$, $\circlearrowleft = \frac{1}{2}(\sigma_0 + \sigma_3)$, and the proposed set of measurements may thus be represented by the two-photon Pauli matrices mentioned earlier; for example, $\rightarrow \otimes \rightarrow = \frac{1}{4}(\sigma_{00} + \sigma_{01} + \sigma_{10} + \sigma_{11})$. The results yield linear combinations of the 2P-SP and may then be inverted. Note that the required measurements are 16, one of which is needed for normalization.

*Degree of two-photon polarization.*— As noted earlier, the degree of polarization distinguishes between pure- and mixed-state beams. A pure state is characterized by $P_1 = 1$. In defining a measure of the degree of two-photon polarization we are faced with additional possibilities for the state: a pure state may further be entangled or separable [12]. The structural form of the 2P-SP reveals information about the entanglement of the state. We examine the case of pure two-photon polarization states in this section and study the case of mixed states in the next section.

If the state is separable and pure, i.e. the two-photons are independent, then the 2P-SP themselves are factorable into the product of 1P-SP, one for each photon, so that $S_{ij} = S_{i0} S_{0j}$. In other words all the two-photon Stokes parameters can be determined through *local measurements*, measurements that are performed on each beam separately. It is easy to show that this always results in the following distribution of values for the 2P-SP:

$$\sum_{j=1}^{3} S_{0j}^2 = 1, \ \sum_{i=1}^{3} S_{i0}^2 = 1, \ \sum_{i,j=1}^{3} S_{ij}^2 = 1. \tag{5}$$

The case of the maximally entangled pure two-photon state [13] leads to

$$S_{01} = S_{02} = S_{03} = 0; \quad S_{10} = S_{20} = S_{30} = 0; \quad \sum_{i,j=1}^{3} S_{ij}^2 = 3, \tag{6}$$

so that each beam, considered separately from the other, is unpolarized, whereas coincidence measurements yield information about the other 2P-SP that describe non-local correlations.

This leads us to a measure of the degree to which the two photons act as a pair and not as two independent photons. We call this the *degree of two-photon polarization* $P_{12}$. This measure coincides with the *degree of entanglement* defined previously [13], which was arrived at via a different rationale. The quantity $P_{12}$ is then given in terms of the 2P-SP by

$$P_{12}^2 = \frac{1}{2}\left(\sum_{i,j=1}^{3} S_{ij}^2 - 1\right) = 1 - \frac{1}{2}\left(\sum_{i=1}^{3} S_{i0}^2 + \sum_{j=1}^{3} S_{0j}^2\right). \tag{7}$$

For the pure state it is clear that this quantity ranges from 1 (maximally entangled state) to 0 (separable state). The (one-photon) degree of polarization, for each photon separately, is $P_1 = \sqrt{S_{10}^2 + S_{20}^2 + S_{30}^2}$ and $P_2 = \sqrt{S_{01}^2 + S_{02}^2 + S_{03}^2}$. One can show that for pure states, $P_1$ is always equal to $P_2$. We note here that there exists a complementarity relationship between $P_{12}$ and $P_j$, namely, $P_{12}^2 + P_j^2 = 1, \ j = 1, 2$ [14]. Since $0 \leq P_1, P_2 \leq 1$, one may then conclude that $0 \leq P_{12} \leq 1$.

*Mixed States.*— The 2P-SP formalism facilitates the study of mixed states as well. We parameterize each state by its degree of two-photon polarization $P_{12}$ and the average one-photon degree of polarization $\overline{P}^2 = \frac{P_1^2 + P_2^2}{2}$ (in general, $P_1 \neq P_2$ for mixed states). There are general constraints on the range of values assumed by $P_{12}$ and $\overline{P}$ for any state.

We plot an outer contour of the possible values of $P_{12}$ and $\bar{P}$ in Fig. 2. The axes are chosen to be $P_{12}^2$ and $\bar{P}^2$ for reasons that will become clear shortly. In the case of mixed states one can have negative values of $P_{12}^2$ according to the definition in Eq. (7). For such a case ($P_{12}^2 < 0$) we define a new parameter $P_m$ defined as

$$P_m^2 = \frac{1}{2}\left(1 - \sum_{i,j=1}^{3} S_{ij}^2\right), \tag{8}$$

such that both $P_{12}^2$ and $P_m^2$ are always $> 0$.

Pure states satisfy the complementarity relationship $P_{12}^2 + \bar{P}^2 = 1$ so that they are represented by points lying along the straight-line segment $\overline{AB}$ in Fig. 2, where $A = \left(P_{12}^2 = 0, \bar{P}^2 = 1\right)$ and $B = \left(P_{12}^2 = 1, \bar{P}^2 = 0\right)$ represent the factorizable and maximally entangled states, respectively. All other points enclosed in the polygon, and on the boundaries other than $\overline{AB}$ in Fig. 2 represent mixed states for which $P_{12}^2 + \bar{P}^2 < 1$.

As an example of a mixed state consider the density matrix

$$\boldsymbol{\rho}_{12} = \lambda |\Psi\rangle\langle\Psi| + \frac{(1-\lambda)}{4}\mathbf{I}_4, \tag{9}$$

which represents the mixture of a pure state $|\Psi\rangle$ (which lies on $\overline{AB}$), and a maximally mixed state represented by the 4×4 identity matrix $\mathbf{I}_4$, with $0 < \lambda < 1$ [15]. Varying the weight $\lambda$ from 1 to 0 moves the point representing the state along a straight line from its location on the pure state locus $\overline{AB}$ ($\lambda = 1$) to the point $D = \left(P_m^2 = 0.5, \bar{P}^2 = 0\right)$, which corresponds to the maximally mixed state ($\lambda = 0$). For a maximally mixed state all the 2P-SP are equal to zero except $S_{00} = 1$.

As a second example of a mixed state consider a mixture of two factorizable pure states

$$\rho_{12} = \lambda |\Psi_1\rangle\langle\Psi_1| + (1-\lambda)|\Psi_2\rangle\langle\Psi_2|, \tag{10}$$

where $|\Psi_1\rangle = |\Psi_A\rangle \otimes |\Psi_B\rangle$ and $|\Psi_2\rangle = |\Psi_C\rangle \otimes |\Psi_D\rangle$; here $|\Psi_A\rangle$, $|\Psi_B\rangle$, $|\Psi_C\rangle$, and $|\Psi_D\rangle$ are one-photon pure states. The shaded area in Fig. 2 represents all the possible states formed by Eq. (10). As more factorizable states are mixed, the shaded area extends toward D. The straight-line segment $\overline{AE}$ represents the mixture in Eq. (10) when $|\Psi_A\rangle = |\Psi_C\rangle$ and $|\Psi_B\rangle$ is orthogonal to $|\Psi_D\rangle$. Point A represents the cases $\lambda = 0$ and $\lambda = 1$, whereas point $E = (P_m^2 = 0.5, \overline{P}^2 = 0.5)$ represents $\lambda = 0.5$ where we have $\rho_{12} = |\Psi_A\rangle\langle\Psi_A| \otimes \frac{1}{2}\mathbf{I}_2$. If we further mix this state with $\rho_{12} = |\Psi_E\rangle\langle\Psi_E| \otimes \frac{1}{2}\mathbf{I}_2$, where $|\Psi_A\rangle$ and the one-photon pure state $|\Psi_E\rangle$ are orthogonal, then the locus of the mixture is the straight-line segment $\overline{DE}$, where the point D is the case of an equally weighted mixture. The point C in the diagram of Fig. 2 represents classically correlated states ($P_1 = P_2 = 0 = P_{12}$). The meaning of negative $P_{12}^2$, and thus the definition of $P_m^2$, now becomes clear: all mixtures of factorizable states lie to the left of the straight-line segment $\overline{AC}$. Although these states are not directly factorizable in the form $\rho_{12} = \rho_1 \otimes \rho_2$ (except those lying on the outer border along the segments $\overline{AE}$ and $\overline{DE}$) they are separable in the sense discussed in Ref. [16]. Their apparent inseparability is due to the mixedness of the state and not from entanglement. The quantity $P_{12}^2$ is defined in Eq. (7) so as to yield negative values for such cases (one may think of inseparability due to mixedness versus entanglement).

*Discussion.* — We have presented a formalism for describing the polarization properties of two-photon states that is an extension of the Stokes-parameters formalism that is well known in classical optics. In contrast to the density-matrix formalism, this extended formalism clearly separates the one- and two-photon characteristics of the state. Moreover, it is useful for visualizing the effect of mixedness on the one- and two-photon characteristics of the state and could serve as a valuable tool for comparing the merits of various purification and concentration protocols that are of current interest in quantum information processing.

This work was supported by the National Science Foundation; by the Center for Subsurface Sensing and Imaging Systems (CenSSIS), an NSF engineering research center; by the David & Lucile Packard Foundation; and by the Defense Advanced Research Projects Agency (DARPA).

# References

[1] G. G. Stokes, Trans. Cambridge Philos. Soc. **9**, 399 (1852) [reprinted in G. G. Stokes, *Mathematical and Physical Papers* (Johnson Reprint Corporation, New York and London, 1966)].

[2] W. A. Shurcliff, *Polarized Light: Production and Use* (Harvard University Press, Cambridge, Massachusetts, 1966); M. Born and E. Wolf, *Principles of Optics* (Cambridge University Press, New York, 1999).

[3] J. Lehner, U. Leonhardt, and H. Paul, Phys. Rev. A **53**, 2727 (1996).

[4] D. N. Klyshko, *Photons and Nonlinear Optics* (Nauka, Moscow, 1980) [translation: Gordon and Breach, New York, 1988]; J. Perina, Z. Hradil, and B. Jurco, *Quantum Optics and Fundamentals of Physics* (Kluwer, Boston, 1994).

[5] D. Bouwmeester, J.-W. Pan, K. Mattle, M. Eibl, H. Weinfurter, and A. Zeilinger, Nature **390**, 575 (1997); D. Boschi, S. Branca, F. De Martini, L. Hardy, and S. Popescu, Phys. Rev. Lett. **80**, 1121 (1998); A. V. Sergienko, M. Atatüre, Z. Walton, G. Jaeger, B. E. A. Saleh, and M. C. Teich, Phys. Rev. A **60**, R2622 (1999).

[6] D. N. Klyshko, Kvantovaya Elektron. (Moscow) **4**, 1056 (1977) [Sov. J. Quantum Electron. **7**, 591 (1977)]; D. Branning, A. L. Migdall, and A. V. Sergienko, Phys. Rev. A **62**, 063808 (2000).

[7] P. Jordan, Z. Phys. **44**, 292 (1927); W. H. McMaster, Rev. Mod. Phys. **33**, 8 (1961); E. L. O'Neill, *Introduction to Statistical Optics* (Addison-Wesley, Reading, Massachusetts, 1963).

[8] W. Pauli, Z. Phys. **43**, 601 (1927); J. J. Sakurai, *Modern Quantum Mechanics* (Addison-Wesley, Reading, Massachusetts, 1994).

[9] What we denote the *two-photon Stokes parameters* is distinct from what is denoted *generalized Stokes parameters* or *double Stokes parameters* in Y. Shi *et al*, Phys. Rev. A **49**, 1999 (1994). The authors of that paper devised a scheme to describe two-photon nonlinear processes, such as two-photon absorption, incoherent frequency doubling, and hyper-Raman scattering, using a Stokes-Mueller-like formalism. It is important to note that in their scheme it is not the optical source that exhibits two-photon-like behavior, but the nonlinear physical process itself that responds to two photons from the incident beam(s). Entanglement is not involved in such processes, and their formalism does not describe the higher-order correlations of the optical beams.

[10] E. Schrödinger, Naturwissenschaften **23**, 807 (1935); **23**, 823 (1935); **23**, 844 (1935) [translation: J. D. Trimmer, Proc. Am. Phil. Soc. **124**, 323 (1980); reprinted in *Quantum Theory and Measurement*, edited by J. A. Wheeler and W. H. Zurek (Princeton University Press, Princeton, NJ, 1983)].

[11] W. K. Wootters, in *Complexity, Entropy, and the Physics of Information: SFI Studies in the Sciences of Complexity*, edited by W. H. Zurek (Addison-Wesley, Reading, Massachusetts, 1990) vol. VIII, pp. 39.

[12] B. E. A. Saleh, A. F. Abouraddy, A. V. Sergienko, and M. C. Teich, Phys. Rev. A **62**, 043816 (2000).

[13] A. F. Abouraddy, A. V. Sergienko, B. E. A. Saleh, and M. C. Teich, "Degree of entanglement for two qubits," submitted to PRA.

[14] G. Jaeger, M. A. Horne, and A. Shimony, Phys. Rev. A **48**, 1023 (1993); G. Jaeger, A. Shimony, and L. Vaidman, Phys. Rev. A **51**, 54 (1995); A. F. Abouraddy, M. B. Nasr, B. E. A. Saleh, A. V. Sergienko, and M. C. Teich, Phys. Rev. A **63**, 063803 (2001).

[15] R. F. Werner, Phys. Rev. A **40**, 4277 (1989); S. Popescu, Phys. Rev. Lett. **74**, 2619 (1995).

[16] A. Peres, Phys. Rev. Lett. **77**, 1413 (1996).

[17] Elements of the approach presented here have been used in connection with electron spin [U. Fano, Rev. Mod. Phys. **55**, 855 (1983)], the investigation of the maximal violation of Bell's inequality for mixed states [R. Horodecki *et al*, Phys. Lett. A **200**, 340 (1995)], and the measurement of the density matrix of a two-photon state [D. F. V. James, P. G. Kwiat, W. J. Munro, and A. G. White, e-print quant-ph/0103121].

**Figure Captions**

Fig. 1. Two-photon polarization state analyzer. S is a two-photon source, $A_1$ and $A_2$ are polarization analyzers, and $D_1$ and $D_2$ are one-photon detectors. $M_1$ and $M_2$ are singles measurements while $M_{12}$ is the coincidence measurement.

Fig. 2. A plot of the possible values for $P_{12}^2$, $P_m^2$, and $\overline{P}^2$.

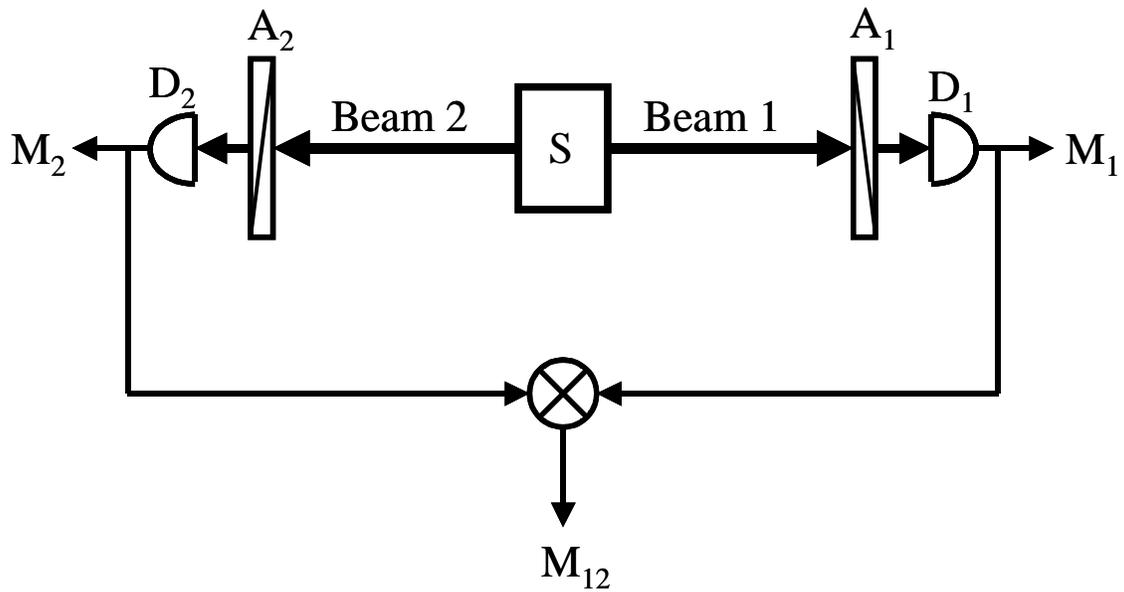

Fig. 1. Two-photon polarization state analyzer. S is a two-photon source, $A_1$ and $A_2$ are polarization analyzers, and $D_1$ and $D_2$ are one-photon detectors. $M_1$ and $M_2$ are singles measurements while $M_{12}$ is the coincidence measurement.

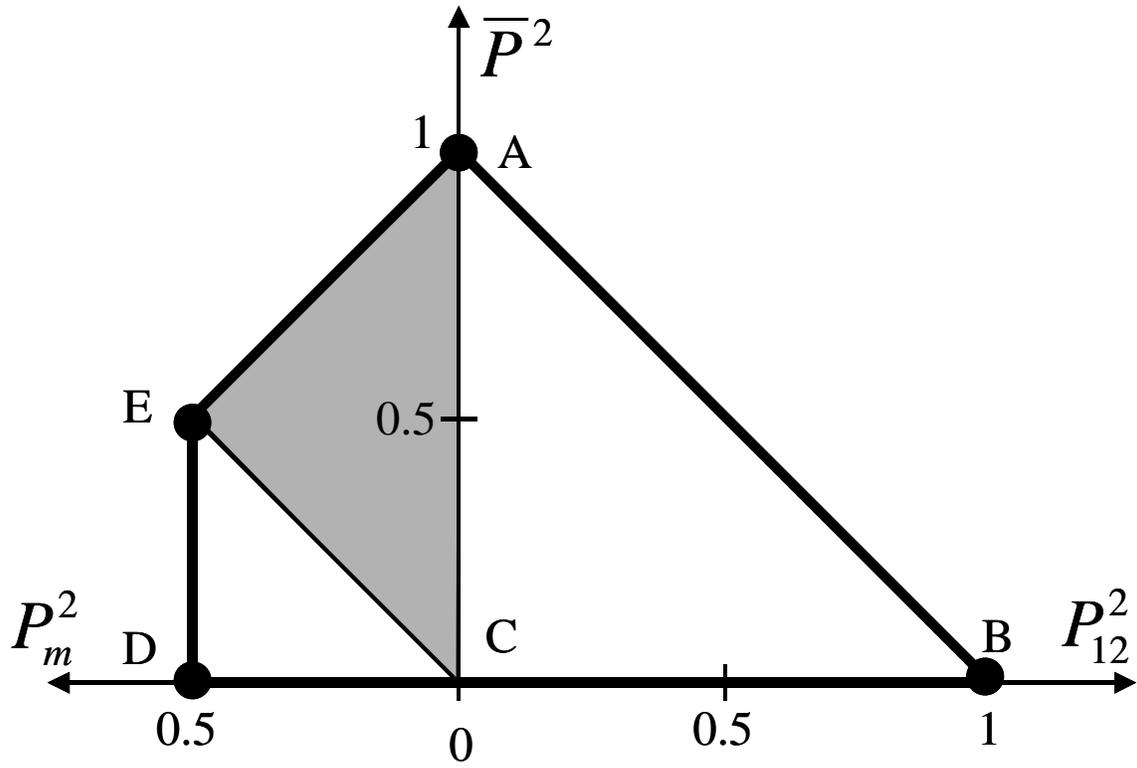

Fig. 2. A plot of the possible values for $P_{12}^2$, $P_m^2$, and $\overline{P}^2$.